\documentclass[times, 10pt,twocolumn]{article}
\usepackage{times}
\usepackage{graphicx}
\usepackage{amsthm}
\usepackage{hyperref}
\usepackage{amssymb}
\usepackage{amsmath}

\begin{document}

\title{ On Framework and Hybrid Auction Approach to the Spectrum Licensing Procedure
      }

\author{Devansh Dikshit ~~~~~~~~ Y. Narahari\\
Electronic Commerce Laboratory, Dept. of Computer Science and Automation, \\
Indian Institute Science, Bangalore,
India. \\Email: \{devansh\},\{hari\}@csa.iisc.ernet.in}

\maketitle
\thispagestyle{empty}
\bibliographystyle{unsrt}

\begin{abstract}
Inspired by the recent developments in the field of Spectrum Auctions, we have tried to provide a comprehensive framework for the complete procedure of Spectrum Licensing. We have identified the various issues the Governments need to decide upon while designing the licensing procedure and what are the various options available in each issue. We also provide an in depth study of how each of this options impact the overall procedure along with theoretical and practical results from the past. Lastly we argue as to how we can combine the positives two most widely used Spectrum Auctions mechanisms into the Hybrid Multiple Round Auction mechanism being proposed by us.
\end{abstract}
\\\\
\textbf{General Terms}\\
Economics, Theory, Spectrum auctions
\\

\section{Introduction}
The world is shrinking. Everyday there is advancement in the field of telecommunications. Telecommunication, which was earlier solely dependent on long cables and then on optical fibres, is now shifting rapidly towards the new wireless technologies, thanks to GSM and its subsequent versions. These technologies are dependent on transmission of airwaves through the atmosphere, the transmission taking place at a certain frequency. The word spectrum defines a range of frequency. The range of frequency can be arbitrarily large, but only a small fraction of it can be used for the telecommunication purposes. The spectrum over a country's area is a property of the people of that country and hence the government of that country. But over past few years, many private parties have entered in the field of telecommunications. And as expected they require a slice of spectrum for that purpose. This also provides a wonderful opportunity for the government as there is useable spectrum which is left unused with the government, with only a fraction of it being used for military and civil purposes by the governments. The amount of this spectrum is usually scarce hence the greater demand-supply gap. Along with generating revenues, privatization of spectrum will also lead to a faster pace of development by the private parties which may not be always possible to do by the government. This has lead to the sale/licensing of the spectrums by various countries across the globe, using methods like lotteries, auctions, tenders etc.

\subsection{Motivation}
Over the past few years, the amount of revenue generated from the spectrum licensing procedure in various countries has far exceeded the government and industrial expectations. With the advent of new technologies, and thus new uses of spectrum, one can expect further rise in the revenue for the future spectrum licences. However when there is big money involved, there is every possibility of attempt by interested parties to rig the procedure, causing a loss in revenue for the state. Apart from the revenue, the government must also ensure the implementation of the social issues involved in the procedure. All these issues, combined with the several market factors, results in a very complex problem of designing the licensing procedure. The market factors we are talking about are the various social, economic, political and geographical features and issues of a country. These factors vary from one country to another and hence it is very difficult to come up with a unified design theory which will work best for all scenarios. Nonetheless, we can always provide a basic framework of the complete auctioning procedure which should be followed by every government going for licensing the spectrum. \\
In our attempt, we have tried to come up with a comprehensive framework which should act as a guideline for the governments in the spectrum licensing procedure. This has been done on the basis of all the research done in this area in the past couple of decades. Since in economics, often the practical results differ from the theory, we have also taken into the account the results of spectrum auctions in various countries in the same period. Essentially, we have separately studied the designing and assigning part of the licensing procedure. As evident from recent results, both are important to the success of a spectrum licensing procedure.\\
On the other hand, the auction of choice of late by the governments across the globe for spectrum licensing has been simultaneous ascending auctions. While this procedure has many advantages over other auction types, a severe drawback is that of possibility of collusive bidding by the bidders. We have designed a new hybrid auction, which takes in the plus points of simultaneous and sequential ascending auctions and at the same time removes drawback of both. We believe that the Hybrid algorithm proposed by us is more robust than any of the existing action mechanism, without compromising on the complexity of the procedure. As will be explained later, this is very important to promote competition.

\subsection{Relevant Work}
There has been a lot of research done in the field of multi-unit auctioning which is applicable to spectrum auctions. Some early pioneering work in this field was done by Milgrom and Weber. \cite{12} Milgrom has another two impact papers in 1989 \cite{15} and 2000 \cite{11} respectively. Both of these deal with the various intricacies of simultaneous ascending auctions. However the spectrum auctions were first used on large scale in FCC auctions at US. The results of these auction provided valuable insights into spectrum auctions and the various strategies employed by the bidders. This is captured by various economists like Milgrom, Cramton, Ausubel, McAfee, McMillan to name a few. All the corresponding research can be found in many publications over the past two decades. \cite{3, 5, 6, 7, 11, 13, 19, 20} is only a representative list of these publications. All these publications are centered over the various issues involved in FCC auctions which was primarily based on simultaneous ascending multiple round auctions. It also provided an insight of the effect of different market factors on the results of the spectrum auction and this forms the basis of the framework of spectrum auctions modeled in this paper. Clock-Proxy auctions introduced by Ausubel, Cramton and Milgrom \cite{1} is another recent development in this area. General multi-item auctions have also been investigated by researchers. \cite{2, 4, 12, 14, 18}\\
As already mentioned, the auction used in majority of the spectrum licensing till date has been simultaneous ascending auction. There are many shortcomings of these auctions which were highlighted time and again. \cite{7, 20, 14} We introduce in this paper a combination of sequential and simultaneous designs of auctions called as Hybrid auctions which we believe should be more practically robust against collusion than simultaneous auctions.

\subsection{Our Contributions}
Our major contributions in this paper are as follows:
\begin{itemize}
\item We have come up with a comprehensive framework of spectrum licensing procedure, based on the recent research and practical results in this field.
\item We have introduced a new Hybrid auction mechanism for spectrum licences which is shown to be more robust with respect to collusive bidding.
\end{itemize}
We will begin by first identifying the ideal targets that a government must keep in mind while initiating the licensing process, in section 2. In this section we will also argue as to why licensing is a better option than selling the spectrum altogether. In section 3, we will investigate what are the various challenges in front of a government while designing the licenses. Then we will move to the basic issue of assigning the licences in section 4. We will also show as to why auctions are preferred over other methods. We will address various issues involved in designing the auction along with a description of their theoretical and practical outcomes. Finally in section 5, we will introduce and analyze the new Hybrid algorithm being proposed by us.

\section{Targets of Spectrum Licensing}
The first step towards the process of licensing should be to identify is the aim of the government in licensing the spectrum. It should identify all the short term as well as long term requirements which must be fulfilled by the licensing process. Some of these targets should be:
\begin{itemize}
\item \textit{Simple and well guarded}: The whole process of licensing should be kept very simple with no ambiguity involved in the guidelines/rules. The process should be simple both from government's perspective (should be easy to conduct), and from the interested player's perspective (should be simple to understand and participate). It should also be well guarded as far as legal issues are concerned.
\item \textit{Promote efficient use of the spectrum}: The final allocation of the spectrum should be in way that the players who have won a share of the spectrum are able to maximize the utilization of the allocated spectrum. This will ensure proper development of technology as well as the community.
\item \textit{Revenue}: Although not the only important issue, it is still one of the most important issues involved, more so in a developing nation like India. The spectrum should not be undervalued in any case. The possibilities of use of spectrum are endless. Nobody would have imagined ten years ago, the multiple ways wireless technologies are being used today. This has been given as the main reason while spectrum auctions have generated much more revenue in recent times than about a decade back. Under no circumstances the government should settle for less than its estimated value. It is better to repeat the licensing process again after some time than to license it at a lower price.
\item \textit{Promote competition and diversity}: As is the case with any trade, the government should aim to avoid monopoly of any player. Putting appropriate caps on the amount of spectrum that can be owned by a single company is one of the many ways to ensure this.
\item \textit{Time taken by the process}: The delay in assigning the licences is itself a cost to the government. As it is India is not one of the early birds as far as 3G licensing is concerned, therefore the whole process should be swift. To give a flavor of the time duration involved, there have been spectrum auctions in US which went on for more than 6 months. The time taken by the process should not be so large that the companies get involved in collusive strategies with either each other or with the auctioneer; at the same time too short on time should not hamper the revenue from the process. The latter case happened during the Wireless Communications Services(WCS) auction at the US in 1997, when the US congress's decision to speed up the process in order to receive revenues for current fiscal lead to a hasty decision and thus poor revenue from the licensing. \cite{20}
\item \textit{Social and Economic Growth}: Proper support and encouragement should also be given to small businesses, women/minority owned firms and to some extent new entrants in order to promote more competition as well as ensuring economic development.
\item \textit{Transparency}: The entire process of licensing should be transparent to all the interested parties involved. This reduces the possibility of corruption and at the same time increases the trust of the interested parties in the government and thus possibly resulting in higher valuation of the spectrum by them.
\end{itemize}

\section{Designing the Licence}
Designing efficient licences is a key to success in spectrum licensing. The proof of design having a pronounced effect on the success of a particular auction design: having five licenses in the UK UMTS auction, rather than four, was critical in stimulating competition and thus extremely high revenues \cite{20}. This was due to the fact that there were 4 incumbents who were favorites to win the auction. The increment of one licence leads to the possibility of a new entrant, while in turn lead to fierce competition. Designing spectrum licences involves complex political, engineering, and economic factors. The various issues and options that a government can face while designing the licences are:
\begin{itemize}
\item \textit{Division of the spectrum (Bandwidth-wise)}: The most important decision is as to how to divide the total available spectrum into various parts. Each of this will have a separate licence. The question is how many parts and of what size. The first part is easier to answer; for this, the government needs to identify the number potential interested parties in a given geographical area, and then the number of licences in that area should be a fraction of that. For the second part, one option is to divide it in equal parts. This approach has its own merit in that it provides flexibility to the bidders during the auctioning process (Will be explained in a later section). But let us suppose we divide the spectrum in a way that there is 1 small licence (bandwidth wise) and rest all are medium sized and identical. A small licence will be obviously of interest to small/new businesses, thus ensuring the goal of economic development. Also this decision can be affected by other issues involved in designing the licence, as discussed below.
\item \textit{Division of the spectrum (Geographical)}: The options available for this issue are whether to have a single set of licences all over the country (as done in UK), or to have different licences for different circles. The first option is suitable for small countries while the latter suits for larger countries. Also anticipating more competition in metros, it is advisable to have more licences in densely populated region. While for the backward/sparsely populated areas, is would be better to have less licences in order to promote competition. In defining the geographic scope of licenses one must consider the ability of the auction mechanism and secondary markets to efficiently aggregate small licenses into larger coverage areas versus the ability of secondary markets to efficiently disaggregate large licenses into smaller areas.
\item \textit{Size}: There are many alternatives to define the size of a license. The reason why this is important is that on the basis of size of a licence only, it is possible to evaluate it both by interested parties as well as the government. Typically size is defined as a combination of following parameters:
    \begin{enumerate}
    \item Bandwidth allotted in the licence
    \item Population of the region where the licence is applicable
    \item Area of the region where the licence is applicable
    \item Literacy rate of the people in the region
    \end{enumerate}
    How these parameters are combined to define the size depends on the government; how much emphasis they are giving to each of these aspects. Generally the first two points are taken into consideration. For example in US spectrum auctions, size is typically measured in MHz-pop: the bandwidth times the population of the region where the licence is applicable.
\item \textit{Duration}: The duration of licences is also a major decision to be taken while designing the licences. With the rapid pace of advancement in communications, there will be many uses of spectrum in future, which might be difficult to visualize now. So a long term licence might undervalue the licensing price of the spectrum in later years. At the same time, it should also be long enough so that the investor has the confidence of recovering his investment. Based on the spectrum auctions taken place in past, duration of around 10 years should be optimal.
\item \textit{Use of Spectrum}: The licence should specifically mention what are ways in which the holder can use the licence, and more importantly what uses are not allowed. There should be a monitoring body for the same. This can be used as a guard against any future use of spectrum which might provide unjust advantage to the holder. Again, this emphasizes the requirement for a strong legal shield for the licence.
\item \textit{Bandwidth Cap}: Another important issue is of the limit on the amount of bandwidth a particular party is allowed to hold in a particular area. This helps in restricting monopolistic trade practices. For example in US auctions, there is a limit of 45 MHz on holdings by a single firm in a particular area. This ensures that there are at least 5 different firms in any given area, even after all the secondary market deals. A variant can be to limit the number of licences a particular firm can hold in any area to 1. This was used in UK spectrum auctions, and was a driving factor behind its success as it negates the collusive strategies between the interested firms. But in the case, that the sale of a part of bandwidth licence is allowed (explained in the next point), this second option is not applicable.
\item \textit{Resale of bandwidth}: In a majority of the spectrum auctions conducted till date, bandwidth cap permitting, the resale of licence was allowed i.e. the firm which won the licence was allowed to sell it in secondary market. This very much helps in stimulating competition as now more players are interested in the licence hoping to earn profit by selling it in secondary market. But in a majority of auctions till date, it was allowed only to sell the full licence, or to divide it geographically. Another option could be to allow the winners to divide the spectrum into bandwidth bands and then sell each band separately. The division should be allowed only at the discrete points (say at multiples of 5 MHz) only. Along with the revenue issue, this will also help in efficiency cause as now with greater flexibility, we can hope that the secondary market will itself remove the inefficiencies of the initial allocation. Again the process of resale should be legally bounded in order to avoid any problems in reauctioning the spectrum once the licence term expires.
\item \textit{Payment Rules}: There are 2 payment options: winner pays the whole amount upfront or he does it in instalments. The second option was used in US auction with the hope that it will stimulate competition by promoting weaker firms. But many of those licences ran into legal tussles with the winner not paying all the instalments.
\item \textit{Coverage of Backward Areas}: Private firms will always be interested in developed areas only, and the state owned firm can not develop all the backward areas on its own. Therefore it is required that the private firms are mandated through the licences that they have to invest in backward area also. Some of the ways to do this:
    \begin{enumerate}
    \item A certain percentage of the population, in the regions in which a firm has the licence, must be rural. This can be made mandatory for all firms above a certain threshold size of the licence.
    \item Firms can be mandated to spend a certain amount (can be possibly a fraction of the cost of the licence) towards the development of communication technology in the region. This should be used with caution as it might result in lowering the revenue.
    \item As mentioned earlier, special help can be given to "designated entries"; state owned/minority owned firms to participate in the licensing process. How to do this? We will answer this question in our next section
    \end{enumerate}
\end{itemize}
\section{Assigning the Licence}
Now we will move on to the second and perhaps the more widely researched part of licensing the spectrum i.e. the process of assigning the spectrum. First we will investigate what are the traditional ways to assign the licence, showing that spectrum auctions should be the preferred way of doing it. Then we will provide a summary on the issues an auctioneer must decide before starting the auction. Finally, we will report about the various types of spectrum auctions methods, citing the advantages and the disadvantages of each method.

\subsection{Methods of Assigning the Licence}
\subsubsection{Administrative Process}
As the name suggests, an administrative process begins with the parties which are interested in the spectrum making a proposal for how they intend to use it and their valuations for the spectrum. The government or the auctioneer, after viewing all the proposals, decides whom to allocate the spectrum. The decision is taken on the basis of all the prospects of the proposal, and the policies of the government. As evident, there are many problems with this procedure: they are very slow, leave scope of corruption and most importantly are not transparent. Thus it is not advisable to go for this approach.
\subsubsection{Lotteries}
In a lottery, the government randomly selects licence winners from the interested firms in a draw of lots. There might be an initial screening process before the lots but the lottery process is transparent to all. As can be easily guessed, the problem with this approach is that a firm with lower valuation for the spectrum might win the licence. Thus the allocation turns out to be inefficient more often than not. Also since it is a game of chance, the players have every incentive to apply in large numbers to improve their chance of winning. Thus we might end up with a single firm applying with say about 50 different proposals. Worse yet, there might be many individuals or small firms or firms which do not have any use of spectrum, applying for the spectrum hoping to gain huge profits by selling it in the secondary market. Thus we might end up getting huge number of proposals for each licence. This results in a waste of resources in processing the applications/proposals. Moreover, the winners are not those best suited to provide a service. It can take years for the licenses to be transferred via private market transactions to those capable of building out a service. \subsubsection{Auctions}
The main advantage of an auction is the fact that it assigns the spectrum to the firm which values it the most along with preserving transparency in the process. Everyone can see why a particular bidder won the auction. Scope of collusion with the government officials is drastically reduced although there are still some other concerns. The revenue generated in the process is also usually high as there is a direct competition between the firms in the auctioning process. Those companies with the highest value for the spectrum likely are willing to bid higher than the others, and hence tend to win the licenses. There are several issues, which are addressed below that can potentially limit the efficiency of spectrum auctions. The success of auctioning very much depends on these decisions regarding the design of the auctions and subtle change in any of these can have a huge impact on the final analysis.
There are also some downside of auctioning process. It only ensures that the bidder with the highest private value wins, not the one with the highest social value. Private and social values can diverge in these auctions because the winners will be competing in a marketplace. Suppose we have 2 licences in a given geographical region and 2 firms A and B bidding for these. Suppose that A already holds one of the licences and that auctioning for the second licence is being organized by the government. Now the social outcome will be that B wins the second licence leading to more competition in the area. But A will no doubt try to beat B in the auctioning process itself in order to establish its monopoly in the area. Thus the private valuation of a might be more than that of B for the second licence. Also, dependence on revenue is usually very high in the auctioning process, although certain designated bidders can be helped by the government.
In spite of all the above drawbacks, the advantages of auction process are proven in spectrum auctions across the world. The advantages of the process easily outweighs the drawbacks many of which can be removed by clever designing of licences and rules. Thus we can safely assume that the best way to assign the licences is through auctioning them. So henceforth, we will refer to the process of assigning the licence as auctioning them and the interested parties as the bidders.

\subsection{Auctioning Issues}
We will now iterate some of the fundamental issues that must be decided while designing an auction. The point to note here is that there is no rule as such which will guarantee the success of the auction. All we can do is theoretically say that what are the options available and how bidders are expected to behave in each of these scenarios. In addition we have the history of spectrum auctions, some successful, some not so, that provide us with an insight into the process and how the bidders try to manipulate the whole process. Also the applicability of each of these vary from auction to auction depending pretty much on the valuation of the licences by the bidders depending upon a large number of issues right from the actually design of licences to the social and economic conditions of the country. So here are some of the basic decisions involved in designing the auction (note that some of these are specific to multiple round auctions):
\subsubsection{Open Multiple vs Single Sealed Bidding}
The first issue in auction design is whether to use a single shot bidding process or to have multiple rounds whereas the bidders get a chance to improve upon their bids. Governments generally choose sealed-bidding process over second option for the simple reason that the process is very simple fast and fuss free. On the other hand, from the bidders prospective, multiple rounds are a better method as it gives them an opportunity to aggressively bid for the spectrum.
The reason for these choices is that the simultaneous sealed-bid auction is more difficult to rig than an ascending auction. This is because a member of a conspiracy to rig the bids can cheat on the conspiracy by submitting a bid secretly; thus sealed bids encourage breakdown of cartels. On the other hand if some member of a cartel goes for cheating in a open ascending bid auction where his bid information is open, he can be immediately punished in the following round by the other members of the cartel by improving upon the bids on the items of that particular members interest.\\
As stated above, the main point of difference between the two processes is that there is much more information available to the bidders in ascending auctions. In an ascending auction, bidders can respond to the behavior of others during the course of the auction. Paul Milgrom and Robert Weber \cite{12}, proved that this increase the average revenue in the auction in a symmetric environment. The ascending auction reveals information about the bidders to the bidders. Revealing information reduces the size of the information rents obtained by bidders, increasing prices on average. This positive relationship between revenues and information transmission during the auction has been labeled "the linkage principle" by Milgrom\cite{15}.\\
Now consider the case where the bidders are asymmetric. Suppose we have one strong bidder and several weak bidders vying for a single licence. Now the strong bidder knows that there is very less competitions against him and if he bids his true valuation, he should win. So his primary target now shifts to profit maximization. He will be searching for minimum bid which should land him the licence. On the other hand the weak bidders know that they are up against a strong bidder. So their best chance of winning is to report their true values. In a single shot sealed bid auction, there is a possibility that the winner is the one of the weak bidders. This will in fact be an inefficient allocation as the valuation of a strong player can be much higher and he lost while trying to maximize his profits. (Generally a strong player is one of the incumbents and incumbents usually have more valuation to spectrum than new entrants). In an ascending auction, the strong bidder would revise its bid upward to win, were it actually the high value bidder.
Ascending auctions also removes the "winners curse" problem. Suppose the government has adopted sealed bid auctioning process. Since the bidders have no information at all about the evaluation of other bidders, it is possible that the winning bid is significantly higher than the second highest bid. Although it might result in better revenue for the government, it will result in depression for the winning firm as he will believe that he has paid much more than the actual cost even if the bid was in fact ex ante optimal. This might result in lesser investment by the winning firm in development of the communication technologies for the Spectrum. Thus it might not be a socially optimal way to auction the spectrum. Although this problem can be solved with the help of second price auctions (Vickey Auctions), it can result in another problem of revenue reduction. This has already happened in some countries where the winning bidder has to pay an amount just over the reserve price.  In an ascending auction, actual competition, rather than an expectation about competition, forces bids to the level achieved, reducing regret. On the other hand this might lead to lesser revenue for the auctioneer although the history suggests otherwise.\\
Sometimes, with the new information revealed during the course of the auctions, the valuations of the bidders may also increase. For example a bidder with a fix budget bidding for two licences might be ready to put more money on second licence when he knows that he might be winning the first below his valuation. Also the firms who are ignorant about certain facts might discover them during the course of the auctioning to revive their valuations and hence their bids. While the weight of the economic literature favors multiple round ascending auctions, the case is far from transparent, primarily because of the history of government auctions using sealed-bids.

\subsubsection{Simultaneous vs Sequential}
The second most important auction design decision is applicable whenever we have multiple inter related objects to be auctioned, prime example being the spectrum auctions. The traditional approach adopted not only by the governments but also by various auction houses across the world (for example Christie's and Sotheby's) is sequential in nature. i.e. at a time bids are accepted only for a single item. There can be multiple rounds as we have in open verbal auctions, but the objects are auctioned in a fixed predetermined order. On the other hand, in a simultaneous auctioning mechanism, the auctioneer takes bids for all the objects simultaneously in each round, spanning over a variable number of rounds determined dynamically during the auctioning process. We will now study pros and cons of each of these designs in detail.\\

Consider the example of sponsored search auctions. Various search engines like Google, Yahoo etc run auctions for determining which sponsored links to display alongside the search results. There are fixed number of slots for the sponsored links. Now consider the case of two slot model, i.e. there are only two slots for sponsored links in some engine. And there are three bidders (A, B, C) bidding for those two slots. Now suppose bidder A's advertisement requires a large space such that it cannot be fitted in a single slot. So bidder A would like to preferably win both the slots so that he can display his full fledged advertisement, otherwise he has to curtail it. On the other hand, B and C are happy with any one of the slots. When the amount a bidder is willing to pay for an item depends on the other items it acquires, sequential auctions deny the bidder crucial information. Suppose now in above example, bidder A values either slot separately at Rs 100 but combined value for both the slots is Rs 300. This bidder on the first slot would be willing to pay up to Rs 200, provided it expected the second slot to sell for no more than Rs 100. On the other hand, having bought the first item, the bidder A would now be willing to pay up to Rs 200 for the second item, even though this creates a loss of Rs 100 on the pair. This problem for the bidder is known as the exposure problem: holding one item exposes one to a loss created by the complementarities in values. The bidder has to forecast the price of future items to bid sensibly on the earlier items. This need to forecast creates a dilemma for the bidder, whether to bid safely and probably lose, or bid aggressively and wind up stuck holding an incomplete aggregation. Only a combinatorial design avoids the exposure problem, but ends up creating other problems, as will be discussed later.\\

Sequential auctions are problematic when items are substitutes as well as complements. Suppose in our above example of sponsored search auctioning, both bidders B and C have a valuation of Rs 150 on either one of the slots. Now suppose bidder A takes the risk and buys the first slot in auction at a price of Rs 151, hoping that he will get the second slot for upto Rs 149. But during the second auction, say bidder B wins with a bid of Rs 150. Thus bidder A end up getting only 1 slot for Rs 151 against his valuation of Rs 100. Thus the process is inefficient in the sense that items are not allocated to the bidder who values them the most. The optimal solution would have been to either give both slots to A or to give one slot each to B and C. Although the revenue generated in this case is more than that in the optimal case, there have been incidences in the past where sequential auctions have led to absurd revenues.\\

The Swiss wireless-local-loop auction conducted in March 2000 illustrates the difficulties of sequential sale. Three nationwide licenses were sold in a sequence of ascending auctions. The first two licenses were for a 28 MHz block; the third was twice as big (56 MHz). Interestingly, the first license sold for 121 million francs, the second for 134 million francs, and the third (the large license) sold for 55 million francs. Empirical auction data have shown such inefficiency to be a problem. In particular Gandal (1997) demonstrated that use of sequential auctions for cable TV franchises in Israel affected revenues, and arguably affected efficiency.\\

In the above example of sponsored search slots, the items to be auctioned were assumed to be identical. In case they differ, there arises another problem with the sequential design, that of ordering the auctions for different items. For example in the scenario of spectrum auctions in India, suppose we have licences for each state separately. Now the problem is to determine the order in which to auction these licences; whether to go from north to south, east to west, smallest to largest etc. Ordering induces a bias; particular firms care more about some markets than others and will prefer learning from the markets they care less about first. In addition, sorting out the most important markets prior to less important complementary markets is advantageous. No ordering is neutral and thus there can be possible spat among the auctioneer and the bidder on the ordering of items. Again it opens up the possibility of collusion of bidders with the government official to auction in an order of their choice.\\

Another serious problem of sequential designs is that it leads to the bidders following a pre specified set of strategies. They are not able to incorporate the information revealed during the auction process into their valuation. This information is mostly regarding the valuations of the other bidders, information about the licences already being sold etc. This happens because of the pace of the auction and the vast amount of money involved. The prime example of this is the recent Indian Premier League's (IPL) player auctions. Teams came for auction with a pre determined set of strategies and the auction ended with highly skilled players getting lesser price than some of not so good players who were auctioned in the later part of the auctioning process. Adoption of pre determined strategies was accepted by teams after the auctioning process. Thus it hampers the basic advantage of having multiple rounds: the information released during the auction process which is expected to increase the valuations of the bidders because it leaves bidders with large decisions in a very limited amount of time.\\

On the other hand, the simultaneous auctions are relatively new entry in the field of auctions. In a simultaneous design, the auctioneer takes the bids for all items simultaneously from the bidders in each round. The auction continues till there is no new bid on any of the items in a round, i.e. the bidding on all the items start and stop simultaneously. The straightforward benefit of this design is that it gives greater flexibility to the bidders who are interested in clubbing the objects i.e. who want to pay more for a particular combination of items. In our previous example of sponsored search auctions, if bidding for both the slots is done simultaneously, player A will come to know during the bidding process about the valuations of bidders B and C for the slots. Since bidding is simultaneous, there is no need for bidder A to take the risk. Although it seems that this might lead to a reduction in revenue (Rs $300$ against Rs $301$), it leads to an efficient allocation of the spectrum because of the flexibility given to the bidders to aggregate the items. Also in case of simultaneous design, the amount of information revealed during the auction is much more than that in case of sequential design.\\

The advantages of simultaneous design over sequential are: firstly now auctioneer don't need to worry about the order of auctioning the items, as all the items are auctioned simultaneously. Secondly, there is no time bound i.e. simultaneous auctions run over several days, so the bidders get enough time to evaluate the information revealed during the auctioning process. This would help them to deviate from pre defined strategies.\\

But the simultaneous design is also not without problems. The problem of collusion was explained above. Along with this another major problem which comes up is that now bidders go for wait and see strategy i.e. they wait for other bidders to bid and then decide upon their own bid. This is in fact the optimal strategy for each bidder.  If all the bidders start following this strategy, there will be no bidding at all.\\

The solution was proposed by Paul Milgrom and Robert Wilson is the activity rule: Each bidder must be active in every round on a specified fraction of the licenses they hope to win. By active it means that he must have either the highest standing bid, or makes a new bid on a particular item. Thus, a bidder that seeks 12 licences would, under a $50\%$ activity rule, be required to be active on at least six licenses. This puts a pressure on the bidders to actively participate in the auctioning process, and at the same time allow them with greater flexibility to substitute to other licences if the one they are bidding on becomes expensive. There are several options also available for the activity rule. In the above mentioned design, the licences are considered to be symmetric i.e. each licence will contribute a single unit towards the activity rule calculation. However it is possible, and is being used in spectrum auction across the world, to have non uniform weights assigned to licences on the basis of size (as calculated from one of the methods described above in licence design), importance, geographical area etc.  Another issue is that of the magnitude of the activity rule. Low activity rule allows bidders to substitute more aggressively. However the substitution by the bidders starts decreasing as the prices of the items go up. So to stop them from adopting wait and see strategy in final rounds of bidding, it is necessary to increase the activity rule. For example in the initial US spectrum auctions, auctioneer used a three phase system: initially a $33\%$ activity requirement, then a $67\%$ requirement, followed by a $100\%$ requirement.\\

Although economists have developed the solution for above mentioned problem, there are still some other problems in simultaneous design for which no guaranteed solution is available till date. The first among this, the collusion among the bidders was explained above. Another problem is that of Demand Reduction which affects both revenue and efficiency of the allocation, and is more pronounced in case where multiple units of similar items are up for auctioning. Take the example of sponsored search auction mentioned above. Suppose the auctioneer has put both the slots for simultaneous auctions. And suppose that now we have only 2 bidders A and B with valuations of Rs 100 for either slot. If both the bidders keep bidding for both the slots, in simultaneous auctions they will end up with one slot each for Rs 100 each. However, suppose each one of them bids for only 1 slot each, then they will win 1 slot each at a price much less than Rs 100. Thus they have incentive to reduce their respective demand. As shown by Ausubel and Cramton (1996), in a multiunit uniform price auctions, generally every equilibrium case is inefficient. Bidders have an incentive to shade their bids for multiple units, and the incentive to shade increases with the quantity being demanded. Hence, large bidders will shade more than small bidders. This differential shading creates inefficiency. The small bidders will tend to inefficiently win licenses that should be won by the large bidders.\\

And lastly, although the revenues received by the governments across the world using SMR auctions have generally exceeded the expectation, in the absence of any benchmark it is still not possible to evaluate the performance of this auction revenue wise. There has been criticism of SMR design over the revenue received and evidence of presence of collusion among the bidders. There is no proof that the revenue received is the highest possible.

\subsubsection{Combinatorial bids}
As mentioned in the earlier sections, in case of auctions of interrelated items like that of spectrum licences, a bidder's value of a license may depend on what other licenses it wins. For example consider the case of sponsored search auction mentioned above. In case of bidder A, he is interested in winning both the slots for advertisement as his advertisement can not fit into a single slot. On the other hand there is one more bidder who wants only a single slot for his advertisement. The value the first bidder places on both the slots combined is Rs 100 while the bidder B has a valuation of Rs 150 for either one of the slots. Now if packaging of bids are not allowed and it is a sequential design auction, than for winning both the slots, bidder A will have to bid for at least Rs 150 in both the rounds. Thus he will end up paying Rs 300, which is Rs 100 more than his valuation. The only equilibrium of this game is that bidder B wins a slot for Rs 150, the other remains unsold. This is both inefficient, as well as less revenue is generated. In case package bids are allowed or simultaneous auctions are performed, bidder A will be able to package the slots and win both of them for Rs 200 which is efficient and revenue maximizing. Note that now the equilibrium of the auction will be bidder A winning both the slots.\\
With a package bid, the bidder either gets the entire combination or nothing. There is no possibility that the bidder will end up winning just some of what it needs. This saves the bidders from the exposure problem in which a bidder might have to pay more than his valuation just because he has already bought a part of the package he is interested in. In the example above, suppose bidder B's valuation for either slots was Rs 100 instead of Rs 150. And suppose bidder A wins the first slot. Then his requirement of 2 slots is exposed to the second bidder, and being his direct competitor, there is incentive for him to continue bidding to increase the price and thus increasing the loss to bidder A. Thus insulation from this problem provides more confidence to the bidders and they bid more aggressively. But package bids leads to many other issues as well.\\

Package bids tend to favor large bidders seeking large aggregations due to a variant of the free-rider problem, called the threshold problem (Bykowsky, et al. 2000, Milgrom 2000). Take the case of sponsored search auction above. Now suppose there is a third bidder C with a valuation of Rs 100 for either slot (Valuation of A = Rs 200 for both slots combined, valuation of B = Rs 150 for either slot). Now the efficient outcome will be both B and C getting a slot each. But suppose in the simultaneous design both start bidding at Rs 75. And bidder A bids Rs 180 for the package. Now since valuations are private information, there is incentive for both B and C to not increase their bids and hope that the other one does it so that the combined bid crosses the Rs 180 bid of bidder A. Thus A would end up winning for Rs 180 which is an inefficient outcome.\\

Another problem with package bids is the complexity of the problem. If all combinations are allowed, even identifying the revenue maximizing assignment is an intractable integer programming problem when there are many bidders and licenses. The problem can be made tractable by restricting the set of allowable combinations (Rothkopf, et al. \cite{22}). But doing so will again lead to the problem of decision of which combinations to allow and which not so that the auction is fair. Again as with the case of order of auctions in sequential auctions, finding fair combinations are not possible. Alternatively, a bid mechanism can be used that puts the computational burden on the bidders. In such system, bidders must propose bids that in combination with other bids exceed the amount bids for standing package bids (Banks, et al. \cite{21}).\\

So the best way out is to allow for resale of licences and hope that the secondary market will remove the inefficiencies in the allocation. Although the revenue received by the government might still not be the maximum possible, it will lead to efficient final allocation. Also allowing resale might improve competition and thus revenue as argued in the previous sections.

\subsubsection{Reserve Price}
The reserve price or the starting price of auction below which licence would not be sold is another tricky issue for the auctioneer to decide. Usually reserve prices are a trade off between revenue and efficiency. If the reserve price is too high the licence might not even get sold. On the other hand keeping low reserve price might lead to less revenue. The point to notice here is that over the past decade, the importance of the spectrum licences has increased many folds. It has become a vital source of revenue for governments over the world. So a compromise on revenue part might not be the greatest idea. On the other hand even in the scenario when due to high reserve price the licence remains unsold with the government, it might be beneficial to auction it again after some time rather than to compromise on the revenue.  A seller who fails to sell has the option of selling in the future; thus the seller's  value may not be the seller's use value, but instead the value of a buyer not yet present; if the seller expects to sell at a higher price in the future, it may be efficient not to sell today. This may be the case when new technologies are possible, technologies that will be resisted by incumbents since the technologies will harm existing services.
Another threat due to low reserve price is that of stock piling by the incumbent. Though it is irrelevant in the case of present spectrum auctions in India, it is worth mentioning that incumbents might have incentive to but the licence not for use but rather than to stock pile them in order to reduce the competition. A low reserve price would certainly help them in this cause.

\subsubsection{Increment Size}
In case of multiple round auctions, another important issue in auction design is: what is the minimum bid increment size allowed between the rounds, over the highest bid of previous round. For example suppose we have 2 bidders A and B bidding for a single item with a reserve price of Rs 100. The private valuations of A and B for the item are Rs 150 and Rs 159 respectively. Now suppose the minimum increment size is of Rs 1. Then the auction will continue for 51 rounds with both bidders increasing their bid by Rs 1 in each round. Note that it is in fact the optimal strategy of bidding. So while we will get the optimal revenue, keeping minimum increment size small will lead to auction being too lengthy and wasting of resources. On the other hand suppose minimum increment size if of Rs 100, then both the bidders will bid only Rs 100 and the item will be allocated according to the tie breaking rule. In this case although the auction is swift, revenue is drastically left. Although we have taken extreme case in this example, it is still clear that keeping minimum increment size large will lead to inefficient allocation.
Now consider the case when minimum bid increment size is Rs 10. Now after 5 rounds, both A and B will bid for Rs 150, and item will again be allocated on the tie breaking rule. Although the auction is swift and revenue loss is also not much, but there is a loss of Rs 9. This highlights the fact that fixed minimum increment size leads are not the optimal solution and that they should be dynamically changed over the course of the auction.
Paul Milgrom came up with an insight that halving the step size or increment approximately doubles the time required to complete the auction. The loss in revenue happens only when the bidder with the second highest value wins because the bidder with the highest value is not willing to pay second highest bid plus the increment. Thus the loss is maximum equal to the increment size. Also the minimum increment size is generally not a absolute value, rather in percentage of previous highest bid. So coupled with the example above, it is easy to see that the best way out is to have high increments in the early rounds when there are more bidders to achieve speed up and low increments in later round to achieve efficiency. Effectively the increment size for a given round should be dependent on the activity (or the number of new bids received representing the number of interested bidders) for that particular licence in the previous round.

\subsubsection{Designated Entries (Preferred Bidders)}
The designated entries are those bidders who are supported by the government for social/political causes. These include state owned firms, businesses owned by minority groups/women. In case of developing country like India, the Indian frims could also be aided by the government by keeping them in this category to provide them level playing field against the foreign giants. There are two ways to support the designated entries. The first one is the traditional and simple way in which certain part of the item, in this case the spectrum is kep aside for the designated entries at a suitable price. But another way of aiding the designated entries, the use of bidder credits, has been developed and used successfully in many spectrum auctions across the world. A bidder credit is similar to the idea of handicapping in sports. Suppose a designated entity, because of lack of capital or expertise, has a $10\%$ lower value on average. Then the auction can level the competition by providing a $10\%$ bidder credit - that is, charging the designated bidder only $90\%$ of their bid. Bidder credits are better compared to the first options due to following reasons:
\begin{itemize}
\item Bidder credits increase competition in the auction. The designated entities become more effective bidders, while the non-designated entities also have a chance to fight for the licenses that would otherwise be set-aside.
\item Bidding credits set a price or value for promotion of designated entities, thereby permitting resale at a cost of refunding the bidder credit.
\item The inefficiency and the revenue loss is now less as the designated entries will get the licences only when their valuations are comparable to market price.
\item Revenues may increase over the levels that would prevail without any bidder credits, because for small value of bidder credit, the increase in competition outweighs the inefficient allocation. On the other hand, revenue will fall as the bidder credit gets larger than the disadvantage of the designated group, and designated groups win majority of the licences. The US spectrum auctions used 25
\item The first policy of keeping aside a part of spectrum for designated entries leads to the creation of small licences by the government to minimize the revenue burden. Also the market players lobby hard to minimize the set aside licences so that their competition is less. In case of bidder credits, larger meaningful licences could be formed.
\item Price preferences naturally apply to partial ownership of the designated entries, by giving partial credit.
    The only problem with bidder credit option is the determination of the size of bidder credit. The point to note here is that bidder credits can be use as an effective handle to differentiate between different designated entries by giving them different amount of credits.
\end{itemize}

\subsubsection{Stopping Criteria and public information}
Another important decision in the auction design is when to stop the auction process in the auctions consisting of multiple rounds over time. The choice is simple in case of sequential auctions when only a single item is auctioned at a time. So if there is no bid in any round the item is allotted to the highest bidder of the previous round. In case of simultaneous auctions, the general trend has been to stop the auction process after a round in which there has been no new bid on any of the items in the fray. But by varying the stopping criteria and slightly changing the design of the auctions, we can overcome the drawbacks of the simultaneous auction while maintaining its strengths. This will be explained in the next section in the hybrid auctions being proposed by us.
Also important is the amount of information to be made public by the auctioneer after each round. In earlier auctions, all the information about the bids including the bidder identity was made public after each round hoping to provide more flexibility to the bidders. But then there were several collusion cases among the bidders. So now-a-days, only the bid amounts are made public and the identity of the bidders is now exposed by the governments.

\subsection{Existing Auction Mechanisms}
We will now describe different auction formats that have been used for spectrum allocation, and discuss the pros and cons of each (See also Agorics, 1996). The last in this list, the Hybrid Ascending Multiple Round (HAMR) auction, is what we are proposing as an alternative to the existing SMR auctions which are being used in quite a few spectrum auctions across the world.

\subsubsection{First-Price Sealed-Bid Auctions}
This is the traditional method of allocating licences in India. Usually done with the help of tenders, this process involves inviting sealed bids from all the interested bidders, and then allocates the licence to the highest bid. Sometimes revenue is not taken as the only criteria but we will not go into those political/social details. The auction is single round, once and for all category auction.\\
There are three main advantages of the first-price sealed-bid auction:
\begin{enumerate}
\item It is simple and easy to conduct.
\item The process of auctioning is fast, which is one of the major advantages in the present case of India.
\item There is no scope of collusion among the bidders as each bidder has to submit sealed closed bids to be opened only after all the bids are submitted. So cartels would break down.
\end{enumerate}
In spite of above advantages and the fact that this has been the preferred choice of auction mechanism over the years, first price sealed bid auction have been long disposed in favor of other mechanisms due to the following reasons:
\begin{enumerate}
\item The auction is not truthful i.e. it is not incentive compatible for the bidder to reveal his true valuation of the licence as his bid. Winning bidder can improve his profit if he bids than his valuation but higher than the bids of all others. Since valuations are private knowledge to all bidders and bids are known only after all the bids are collected, each bidder relies on speculation/past experience to speculate about the bids of other players and then might shade or lower his bid accordingly.
\item Since bidders tend to shade their bids, there is a strong possibility that the bidder who wins the licence might not be the one who values it the most i.e. the final allocation is not efficient.
\item First price sealed bid auctions also suffer from the winners curse problem, which was described above in the auction design issues section. In some cases, the winner might actually go bankrupt (for example, NextWave Communications in the U.S.) by borrowing a lot to bid high, but by not being able to service this debt in the loan period through revenue generation. This is a socially unacceptable scenario.
\end{enumerate}
Note that the first-price sealed-bid auction is similar to what is called a "Dutch auction". In a Dutch auction, used to sell tulips, the auctioneer starts with a given (usually high) price for the resource and then progressively lowers its asking price. The person who cries out the first bid for a resource gets it; the one with the highest or first price gets the resource being auctioned.

\subsubsection{Sealed-Bid Vickrey Auctions}
The famous Vickery second price sealed bid auctions were invented by William Vickrey in 1961. Vickery auctions are truthful i.e. it is dominating strategy for each bidder to report his true valuations as his bid. Sealed-bid Vickrey auctions are very similar to the first price sealed bid auction with the only difference being that now the winner has to pay an amount equal to the second highest bid. Thus changing only the payment rule, results in the auction becoming truthful in nature. Like the first-price auction, the bids are sealed, and each bidder is ignorant of other bids unless there is collusion. The item is awarded to highest bidder at a price equal to the second-highest bid (or highest unsuccessful bid). In other words, a winner pays less than the highest bid. If, for example, bidder A bids Rs 10 and B bids Rs 15 and C offers Rs 20, bidder C would win, however he would only pay the price of the second-highest bid, namely Rs 15.\\
The advantages of this auction design over the first price auctions are:
\begin{enumerate}
\item Vickrey auctions are both economically efficient and truth revealing. It is easy to show that, because the winning bidder needs only pay the second highest bid, there is no incentive to "cheat" and misrepresent the true value of the resource, and thus the item is won by the bidder who actually values it the most.
\item Vickery auctions don not suffer from the winners curse problem, as the winner has to pay an amount equal to the second highest bid.
\item There is no dip in the revenue. Note that although bidders bid their true valuations, the winner pays amount equal to the second highest bid. Thus one might expect that there is a dip in revenue when compared to the first price sealed bid auction. Nobel Prize winning economist Roger Myerson Proved in 1981 the famous revenue equivalence theorem which basically proves that the expected revenue under both a sealed-bid first-price auction and a Vickrey auction (sealed-bid or English) are the same.
\end{enumerate}
Apart from these reasons, Vickery auctions are better when compared to multiple round auctions for the same reasons the first price auction was. The main advantage of both of the above designs over the open bid-multiple round designs is the fact that they are much more resistant to collusion by the bidders. It is also very difficult to rig the process of auctioning, apart from lobbying for the sequence in which the licences are to be auctioned.\\

The downside of the Vickery auctions is that in spite of the revenue equivalancec theorem, in practice, it might happen that revenue generated is not of the liking of the government. The second bid might be ridiculously lower than the first bid, as happened in case of spectrum auction in New Zealand where the winning bid was much higher but the second highest bid was just above the reserve price. So the licence was sold just above the reserve price.\\

The general problem which exists in both the above auctions is that the bidders might not even know their own maximum valuations of the licence, i.e. after having a flavour of other bidders' valuations; they might be willing to improve their bids in light of new information. This is specific to Spectrum Auctions in the sense that first it is a new and developing technology market, so it's very difficult to predict the utility of the licence by the bidder. The problem is similar to the one faced by the auctioneer while deciding upon the expected price or base price. Secondly, all the licences are inter-related. Since the bidders might be interested more in packages (as explained earlier in combinatorial bids) of licences rather than each individual licence, they might not be able to evaluate each one optimally, separately, and their valuation might be substantially greater with the new information than without any prior information.

\subsubsection{Simultaneous/Sequential Ascending Multiple Round Auctions}
The solution to above problems led to the use of sequential ascending multiple round auctions. This is precisely the way individual items are auctioned by the auction houses. The player's auction, in the recently concluded Indian Premiere League, is another example. Sequential auctions are conducted for each item separately. Each auction starts with a base price and then in each round, auctioneer calls for new bids from the bidders in accordance with the increment rule. The rounds continue till there is no new bid by any bidder. The item is then allocated to the bidder with the current highest bid.\\

In the simultaneous design, the bids for all the items, in this case the licences are called simultaneously from all the bidders. The auction can continue for many days as now bidders do have the option to try and form favorable packages. The rounds stop when there is no new bid for any of the items. Thus the process starts and stops for all the items simultaneously.  Other aspects of these auctions were discussed in details while explaining the auction design issues above and it was also argued why the simultaneous design is better than the sequential. In fact, simultaneous design has been the choice for various governments for quite some time now. Here we will discuss the advantages and disadvantages of simultaneous ascending multiple round auctions.\\

The advantages of Simultaneous Ascending Multiple Round (SAMR) auctions, over the previous two auctions discussed above are:
\begin{enumerate}
\item Having multiple rounds of auctions results in efficient allocation.
\item There is no winners curse as the winner has to pay second highest bid plus an increment.
\item Much more information available to bidders during auctions which might act as a favorable catalyst for revenue generated.
\item Bidders tend to re-evaluate their bids over the course of auction, and the long duration of auction provides them ample time to take the new information revealed during the previous rounds into account and make major decision over their future bids.
\end{enumerate}
However, SAMR does have some drawbacks also. These have already been discussed in various sections above, but to summarize we have the following disadvantages:
\begin{enumerate}
\item The whole method is complex.
\item The process can be too lengthy when compared to single round auctions.
\item Although the revenue generated by using SAMR auctions has usually exceeded expectations, there is no theoretical proof that SAMR auction are revenue maximization auctions.
\item Collusion among the bidders has been a major problem for SAMR spectrum auctions across the world.
\item Demand reduction also becomes a problem in case a firm is allowed to have more than 1 licence in a particular geographical region.
\end{enumerate}
Thus the SAMR auction also has not been the perfect auction design. Many other, even more complex, designs have been proposed by various economists (for example Clock-Proxy auction by Cramton et.al.) but we will not explore those here.

\section{Hybrid Ascending Multiple Round Auction}
For sake of simplicity we will call drop "ascending multiple round" from all the three (sequential, simultaneous and hybrid) auction mechanisms. We have seen the logic behind using multiple rounds for spectrum auctions. But both simultaneous and sequential multiple round auctions have their pros and cons. The biggest advantage of using a sequential design is that it virtually removes the possibility of collusion despite using open bids. There has been a lot of evidence in spectrum auctioning history where using Simultaneous auctions have led to collusion among the bidders and hence significantly lowering of the revenue. On the other hand numerous disadvantages of the sequential design were also listed above.\\
We propose Hybrid multiple round ascending auction, a novel auctioning mechanism which combines the plus points of both the above designs. In addition to changing the order of rounds, it also has unique stopping criteria in form of \textbf{saturation factor}. This is done in order to speed up the process and also to stop it from degenerating into simultaneous auction. We will first explain the design and then argue about its usefulness.\\

\subsection{Design}
Consider for example the scenario in which we have $3$ licences A,B and C for sale. Then, the sequential auction will have first all the bidding rounds of A (assuming that the order is A,B,C), followed by the bidding round of B, and finally those of C. In case of simultaneous, in each round bidders will simultaneous bid on all the 4 licences till there is no further new bid on any of those. In our proposed hybrid design, in each round of bidding, bidders will sequentially bid on the licences. For example in above case, bidders will first bid for A (single round), then for B, then C and then D. In the next cycle, again each of these licences will have one more round of bidding sequentially. The difference is illustrated in the picture below.\\

The basic goal which we are trying to achieve here is to make available more information to the bidders than in sequential, and at the same time ending bidding of different items at different ties. This will effectively reduce the possibility of collusive bidding. Recall that in simultaneous auctions, the bidding was over when there was no new bid on any item in any particular round. Note that if we use the same stopping criteria in the hybrid auction, that we used in simultaneous, then hybrid will also degenerate into simultaneous in the final analysis.\\
\textbf{saturation factor}: To overcome the above issue, in our design we propose the use of a \textit{saturation factor} (SF) for each of the licences which will control the stopping of bidding for that particular licence. This is a kind of reverse activity rule that is activity rule on items rather than on bidders. SF also helps in avoiding collusion as will be explained later. The SF is initially set to $0$ for every licence. In every round when there is no new bid on licence $i$, $SF_i$ is increased by $1$. Bidding on licence $i$ stops when $SF_i$ reaches a 'threshold saturation factor' $TSF_i$. This threshold can be used as a controlling factor by the government to speed up the whole process. Lower threshold will mean that the auctioning on that particular item will end briskly. On the other hand, government might want to set a higher threshold for important/lucrative licences in order to provide bidders with more time to process the information gained during the auctioning process, hoping that it would ultimately result in higher revenues. We can easily arrive at the following result:\\

\noindent\textbf{Claim $1$}: Collusive bidding can not be an equilibrium strategy for the bidders.\\
\textit{Proof}: The proof is based on the observation that although bidding for all the licences goes on simultaneously, bidding for no two items can end simultaneously. Suppose 2 bidders, X and Y, decide to collude upon 2 licences, say A and C in above example, with X getting A and Y getting C. Also lets us assume without the loss of generality that the order of bidding is the same as in above that is bids for A are taken before those for C. Suppose that the threshold for both the licences is same $(TSF_A = TSF_C)$, and till the round when the saturation factor for both A and C is just 1 less than the $TSF$, both X and Y are colluding not to increase the bids. Then in this case, in the next round if Y does not increase the bid on A (abiding the collusion), then it has the danger of X increasing the bid on C. This is possible because by that time, X would have already won the licence A. Thus the best strategy for X would be to break the cartel and increase the bid on C as Y is his competitor in the market. Thus Hybrid helps in avoiding the collusion just like the way sequential design does.\\

Hybrid reveals much more information about the valuation of the licences by the different bidders than the sequential. But due to its design, it also suffers from some of the drawbacks of the sequential auctions. The major among the is the ordering of licences for auctioning. This can be avoided by randomly selecting orders in each rounds. This could not have been done in sequential auction. Another point which should be mentioned here is that the saturation factor (SF) also helps in bounding the time required for the auctioning process. Normally the simultaneous auction when used in different spectrum auctions ran into many months. Sequential bidding in each round might result in even more elongated process. But by using saturation factor, we should be in fact able to speed up the whole process. Its use is analogous to the use of activity rule for bidders, which prompts them to bid on more and more licences. By saturation factor we will be able to pull bidders to licences as when SF on any item reaches near its TSF, bidders will know that bidding on that licence is about to end and it's now or never for them. Of course, the TSF and current SF will be common knowledge to all the bidders.

\section{Future Directions}
There are several research directions possible in this area:
\begin{itemize}
\item Till date, there is no model to measure the revenue efficiency of multiple round auctions. It will be an interesting direction to come up with such a model and then to evaluate various different multiple round auctions explained above, on that model.
\item One can investigate the effect of ordering of items on the total revenue generated both in case of simultaneous and sequential auctions.
\item A mathematical model to characterize various market factors and their effect on the spectrum licensing procedure would be the ultimate target for complete understanding of the process. It can easily be a highly complex model so it will require a lot of market investigation and research
\end{itemize}
In this paper, we have tried to come up with a comprehensive framework for spectrum auctions based on current literature. We expect this to act as a guide for future spectrum licensing procedures especially for those countries which are new in this field. We also introduced Hybrid auction which we believe should be a practical replacement for the existing simultaneous multiple round auctions.

\end{document}